\RequirePackage{fix-cm}
\documentclass[referee, 12pt]{svjour3}       
\smartqed  
\usepackage{amsmath}
\usepackage{amssymb}
\usepackage{bm}
\usepackage{graphicx}
\usepackage{subfigure}
\usepackage{mathtools}
\usepackage{enumerate}
\usepackage{algorithm}
\usepackage{algpseudocode}
\usepackage{setspace}
\usepackage{setspace,xspace}
\usepackage{xcolor}
\usepackage[sort]{natbib}
\setcitestyle{aysep={}}

\usepackage{lineno}
\usepackage{endfloat}

\DeclareMathAlphabet      {\mathbf}{OT1}{cmr}{bx}{n}

\newcommand{\mpr}{\textcolor{blue}{MPR}\xspace}

\newcommand{\bfs}{\mathbf{s}}

\journalname{Mathematical Geoscience}
\begin{document}

\title{GPU-accelerated Simulation of Massive Spatial Data based on the Modified Planar Rotator Model
}
\titlerunning{GPU-accelerated Simulation of MPR Model
}
\author{Milan \v{Z}ukovi\v{c}         \and
        Michal Borovsk\'{y}         \and
				Mat\'{u}\v{s} Lach         \and
				Dionissios T. Hristopulos
}


\institute{Milan \v{Z}ukovi\v{c} \at
              Institute of Physics, Faculty of Science, P. J. \v{S}af\'arik University, Park Angelinum 9, 041 54 Ko\v{s}ice, Slovakia \\
              Tel.: +421-552342544\\
              \email{milan.zukovic@upjs.sk} 
           \and
              Michal Borovsk\'{y} \at
              Institute of Physics, Faculty of Science, P. J. \v{S}af\'arik University, Park Angelinum 9, 041 54 Ko\v{s}ice, Slovakia \\
              Tel.: +421-552342544\\
              \email{michal.borovsky@upjs.sk} 
           \and
					    Mat\'{u}\v{s} Lach \at
              Institute of Physics, Faculty of Science, P. J. \v{S}af\'arik University, Park Angelinum 9, 041 54 Ko\v{s}ice, Slovakia \\
              Tel.: +421-552342565\\
              \email{matus.lach@student.upjs.sk} 
           \and
							Dionissios T. Hristopulos \at
              Geostatistics Laboratory, School of Mineral Resources Engineering, Technical University of Crete, Chania 73100, Greece \\
              Tel.: +30-28210-37688\\
              Fax: +30-28210-37853\\
              \email{dionisi@mred.tuc.gr} 
}

\date{Received: date / Accepted: date}

\maketitle

\begin{abstract}
A novel Gibbs Markov random field for spatial data on Cartesian grids based on the modified planar rotator (\mpr) model of statistical physics has been recently introduced for efficient and automatic interpolation of big data sets, such as satellite and radar images. The \mpr model does not rely on Gaussian assumptions. Spatial correlations are captured via nearest-neighbor interactions between transformed variables. This allows vectorization of the model which, along with an efficient hybrid Monte Carlo algorithm, leads to fast execution times that scale approximately linearly with system size. The present study takes advantage of the short-range nature of the interactions between the \mpr variables  to parallelize the algorithm on graphics processing units (GPU) in the Compute Unified Device Architecture (CUDA) programming environment. It is shown that, for the processors employed, the GPU implementation can lead to impressive computational speedups, up to almost 500 times on large grids, compared to the single-processor calculations. Consequently, massive data sets comprising  millions of data points can be automatically processed in less than one second on an ordinary GPU.

\keywords{Spatial interpolation \and Hybrid Monte Carlo \and Non-Gaussian model \and Conditional simulation \and GPU parallel computing \and CUDA}
\end{abstract}

\section{Introduction}
\label{sec:introduction}
As massive remotely sensed spatio-temporal datasets are becoming more and more common, scalable statistical methods are needed for their efficient (preferably real time) processing. In particular, the data often include gaps that need to be filled to obtain continuous maps of observed variables so that  environmental managers can make prompt and informed decisions and in order to avoid the adverse missing-data impact on statistical estimates of means and trends~\citep{Sickles07}.  Typical cases of gap-filling problems involve remote-sensing images with missing data due to sensor malfunctions or cloud coverage and faulty data that lead to partially sampled grids. In addition, resampling via interpolation is needed in order to combine information from satellite products with different spatial resolution~\citep{Stein02,Poggio12,Atkinson14}. Another problem of interest is the filling of gaps in gridded digital terrain models~\citep{Katzil00}.  The gap-filling can be accomplished using various spatial interpolation techniques. However, most of traditional methods, such as kriging~\citep{wack03}, are computationally too expensive to handle such an ever-increasing amount of data. More recently several modifications have been implemented~\citep{furr06,cres08,hart08,kauf08,ingram08,zhong16,marco18}, which have increased the computational efficiency of kriging-based methods.

Some alternative approaches inspired from statistical mechanics have been proposed  to alleviate the computational burden of kriging methods. Models based on Boltzmann-Gibbs exponential joint densities that capture spatial correlations by means of short-range interactions instead of the experimental variogram~\citep{dth03,dthsel07,dth15}, are computationally efficient and applicable to both gridded and scattered Gaussian data. To also overcome the restrictive joint Gaussian assumption the concept was further extended to non-Gaussian gridded data by means of non-parametric models based on classical spin models~\citep{mz-dth09a,mz-dth09b} and generalizations that involve geometric constraints~\citep{mz-dth13a,mz-dth13b}.

Nevertheless, as long as the computations are performed  in a serial manner, the above methods fail to fully exploit  the available computational resources. With new developments in hardware architecture, another possibility to overcome the computational inefficiency is provided by parallel computation. Multi-core CPU and GPU hardware architectures have now become a standard even in common personal computers.  Thus, the parallel implementations of various interpolation methods is a relatively cheap method to boost their efficiency. So far, several parallel implementations were applied to the most common interpolation methods, such as kriging and inverse distance weighting (IDW), on high performance and distributed architectures~\citep{kerry98,cheng10,guan10,pesq11,hu15} and general-purpose computing on graphics processing units (GPU)~\citep{xia11,tahma12,cheng13,guti14,mei14,stoj14,mei17,zhang18,zhan18}. These studies have demonstrated that  significant computational speedups, up to nearly two orders of magnitude, can be achieved over traditional single CPU calculations by means of parallelization.

The present paper demonstrates the benefits of parallel implementation on GPU of the recently introduced gap filling method that is based on the \emph{modified planar rotator} (\mpr) model~\citep{mz-dth18}. The \mpr model has been shown to be competitive with respect to several other interpolation methods in terms of the prediction performance. It is also promising for the automated processing of large data sets sampled on regular spatial grids, typical in remote sensing, due to its computational efficiency and ability to perform without user intervention. In the \mpr model,  parallelization is feasible due to the short-range (nearest-neighbor)  interactions between the \mpr variables (spins). Recent developments in spin model simulations~\citep{weig11,weig12} have demonstrated that significant speedups can be achieved by using a highly parallel architecture of GPUs. The present paper shows that, for the CPU and GPU used in the tests, the GPU implementation can lead to enormous computational speedup, almost by 500 times  compared to single-processor calculations, for massive data sets that involve millions of data points.

The remainder of the manuscript is structured as follows: Section~\ref{sec:mpr} presents a brief overview of the \mpr model and parameter estimation; more details are given in~\citet{mz-dth18}. In Sect.~\ref{sec:cuda} the CUDA (Compute Unified Device Architecture)
implementation of the \mpr model is presented. The statistical and computational performance of the model is
investigated in Sect.~\ref{sec:results}. Finally, Sect.~\ref{sec:conclusion} presents a summary and conclusions.

\section{MPR Method for Gap Filling}
\label{sec:mpr}

\subsection{Problem definition}
First,  the gap-filling problem to be addressed is defined. A partially sampled two-dimensional square grid $\mathcal{G}$ is considered with $L$ nodes per side.
The sample set involves the grid nodes $\mathcal{G}_{S} = \{ \bfs_{n} \}_{n=1}^{N}$, where $N < L^2$ and the sample values $\mathbf{Z}_{s} = (z_{1}, \ldots, z_{N})^\top$ of the
spatial process $Z(\bfs)$ (where $\top$ denotes the matrix transpose). The goal is to obtain estimates $\mathbf{\hat{Z}}_{p} = (\hat{z}_{1}, \ldots, \hat{z}_{P})^\top$
of the process at the grid nodes $\mathcal{G}_{P} = \{ \tilde{\bfs}_{p} \}_{n=1}^{P}$ where the data are missing. The full grid is obtained by the union of $\mathcal{G}_{P}$ and $\mathcal{G}_{S}$, i.e, $\mathcal{G}= \mathcal{G}_{P} \cup \mathcal{G}_{S}$. No assumptions about the probability distribution of the spatial process are made. On the other hand, it is assumed that the spatial correlations are imposed by means of local interactions between the nodes of $\mathcal{G}$. This assumption  is common in statistical physics and  also underlies the concept of conditional independence. The latter is fundamental in the theory of Gaussian Markov random fields and states that the value at any grid node, conditionally on the values within a small neighborhood around it, is independent of
other grid nodes~\citep{Rue05}. The local interaction assumption has proved adequate for modeling spatial processes that do not involve long-range correlations~\citep{mz-dth09a,mz-dth09b,mz-dth13a,mz-dth13b,dth15,mz-dth18}.

\subsection{MPR model}
This section focuses on an efficient and automatic simulation method (hereafter called the \mpr method), that has been recently introduced
for the prediction of non-Gaussian data partially sampled on Cartesian grids~\citep{mz-dth18}. The \mpr method is
based on a Gibbs-Markov random field (GMRF) that employs the modified planar rotator (\mpr) model.
The idea of the \mpr method is to transform the original data to continuously-valued ``spin'' variables by mapping from the original space $V$ to the spin angle space $[0, 2\pi]$ (assuming ergodicity so that the data sample the entire space $V$),  using the linear transformation

\begin{equation}
\label{map}
\mathbf{Z}_{s} \mapsto \mathbf{\Phi}_{s} = \frac{2\pi(\mathbf{Z}_{s}- z_{s,\min})}{(z_{s,\max} - z_{s,\min})},
\end{equation}
where $z_{s,\min}$ and $z_{s,\max}$ are the minimum and maximum sample values and $\mathbf{\Phi}_{s}=\{\phi_{i}\}_{i=1}^{N}$ and $\phi_{i} \in [0,2\pi]$, for $i=1, \ldots, N$.

In the \mpr model, spatial correlations are captured via short-range interactions between the spins using a modified version of the well known planar rotator model from statistical physics. This approach can account for spatial correlations that are typical in geophysical and environmental data sets. The interactions implicitly determine the
covariance function. The energy functional ${\mathcal H}$ of the \mpr model measures the ``cost'' of each spatial configuration: higher-cost configurations have a lower probability of occurrence than lower-cost ones. The \mpr energy function is given by
\begin{equation}
\label{Hamiltonian_mod}
{\mathcal H}=-J\sum_{\langle i,j \rangle}\cos[q(\phi_i-\phi_j)].
\end{equation}
In equation~\eqref{Hamiltonian_mod} the exchange interaction parameter  $J>0$  tends to favor positive values of the cosine, since they lead to negative energy contributions. The symbol $\langle i,j \rangle$ denotes the sum over nearest neighbor spins on the grid. On the square grid the nearest neighbors of each site involve four spins (left, right, top and bottom neighbors).  The upper bound of $q$ is set to $1/2$ so that the values of $q(\phi_i-\phi_j)$ are restricted within $[-\pi, \pi]$. Finally, $q \leq 1/2$ is the coupling parameter. Smaller values of $q$ reduce large spin contrasts $\phi_i-\phi_j$, while larger values of $q$ tend to increase the contrast. In principle, $q$ can be learned from the data. However, herein its value is set to $q=1/2$.

The joint probability density function of the GMRF is then given by
\begin{equation}
 \label{eq:bg-pdf}
 f =\frac{1}{{\mathcal Z}} \exp(-{\mathcal H}/k_{B}T),
\end{equation}
where the normalization constant ${{\mathcal Z}}$ is the partition function, $k_B$ is the Boltzmann constant, and $T$ is the temperature parameter (higher temperature favors larger fluctuation variance).  The temperature can absorb both the exchange interaction parameter and the Boltzmann constant. Hence,  it is measured in dimensionless units~\citep{mz-dth_2018} and it is the only model parameter (since $q$ is fixed).

\subsection{Parameter estimation and simulation at unmeasured sites}
Assuming ergodic conditions, the temperature is automatically and efficiently estimated using the \emph{specific energy matching principle}, which is analogous to the classical statistical method of moments~\citep{mz-dth_2018}. In particular, the sample \mpr specific energy  is given by
\begin{equation}
\label{eq:mpr-sse}
e_s = - \frac{1}{N_{SP}}\sum_{i = 1}^{N}\sum_{j \in nn(i)}\cos[q(\phi_i-\phi_j)],
\end{equation}
where $j \in nn(i)$ denotes the sum over the non-missing nearest neighbors of the site $\bfs_{i}$ (i.e., $\bfs_{j} \in \mathcal{G}_{S}$), and $N_{SP}$ represents the total number of the nearest-neighbor sample pairs.
The sample specific energy is matched with the equilibrium \mpr specific energy given by
\begin{equation}
\label{ene}
e(T,L) = \frac{\langle {\mathcal H} \rangle}{N_{GP}},
\end{equation}
where $\langle {\mathcal H} \rangle$ is the expectation of the \mpr energy over all probable spin states on $\mathcal{G}$, and $N_{GP}=2L(L-1)$ is the number of nearest-neighbor pairs on the $L \times L$ grid with open boundary conditions. The value of $\langle {\mathcal H} \rangle$ is determined by running unconditional \emph{Markov Chain Monte Carlo} (MCMC) simulations of the \mpr model and averaging the energy over the states of the simulation ensemble. In the unconditional simulations the values of the spins can be varied at every grid node.
The simulation is performed for different temperature values keeping the grid size $L$ fixed.
The value of the sample specific energy $e_{s}$ is estimated directly from the data. Finally, the characteristic temperature of the gappy sample $\hat{T}$ is obtained by
\[
\hat{T} = \arg\min_{T} \lVert e_{s} -  e(T,L)  \rVert.
\]

Once the optimal temperature has been determined, the spatial prediction at the missing data sites $\mathcal{G}_{P}$ is based on \emph{conditional MCMC} simulations
during which the existing sample values  in $\mathcal{G}_{S}$ are kept fixed. The prediction is finally given by the mean of the resulting conditional distribution at the target site based on the simulations.
In thermodynamic equilibrium, the \mpr model has been shown to display a flexible  correlation structure controlled by the temperature.

\begin{algorithm}[t!]
\caption{Hybrid updating algorithm that combines deterministic over-relaxation with the stochastic Metropolis step. $\mathbf{\hat{\Phi}}^{\mathrm{old}}$ is the
initial spin state, and $\mathbf{\hat{\Phi}}^{\mathrm{new}}$ is the new spin state. $\mathbf{\hat{\Phi}}^{\mathrm{old}}_{-p}$ is the initial spin
state excluding the point labeled by $p$. $U(0,1)$ denotes  the uniform probability distribution in $[0, 1]$.}
\label{algo:mpr-relax}
\begin{algorithmic}
\Procedure{Update}{$\mathbf{\hat{\Phi}}^{\mathrm{new}},\mathbf{\hat{\Phi}}^{\mathrm{old}},a,\hat{T}$}
\For{$p=1, \ldots, P$}  \Comment Loop over prediction sites
\State 1: ${\hat{\Phi}'}_p \gets \mathcal{R}\{\hat{\Phi}_p^{\mathrm{old}} \}$ \Comment Over-relaxation step according to~\eqref{eq:over-relax}
\State 2:   $u \gets U(0,1)$ \Comment Generate uniform random number
\State 3:   ${\hat{\Phi}''}_p \gets {\hat{\Phi}'}_p + 2\pi (u -0.5)/a \pmod{2\pi}$ \Comment Propose spin update
\State 4:   $ \Delta \mathcal{H} = \mathcal{H}({\hat{\Phi}''}_p, \mathbf{\hat{\Phi}}^{\mathrm{old}}_{-p}) - \mathcal{H}({\hat{\Phi}'}_p, \mathbf{\hat{\Phi}}^{\mathrm{old}}_{-p})$ \Comment Calculate energy change
\State 5:  $AP = \min\{1,\exp(-\Delta \mathcal{H}/\hat{T})\}$ \Comment Calculate acceptance probability
\State 6:  $\mathbf{\hat{\Phi}}^{\mathrm{new}}_{-p} \gets \mathbf{\hat{\Phi}}^{\mathrm{old}}_{-p}$ \Comment Perform Metropolis update
            \If{$AP > r \gets U(0,1)$}
            \State 6.1: {$\hat{\Phi}_p^{\mathrm{new}} \gets {\hat{\Phi}''}_p$}    \Comment Update the state
            \Else
            \State 6.2: {$\hat{\Phi}_p^{\mathrm{new}} \gets {\hat{\Phi}'}_p$} \Comment Keep the current state
\EndIf
\EndFor \Comment End of prediction loop
\State 7: \Return $\mathbf{\hat{\Phi}}^{\mathrm{new}}$  \Comment Return the ``updated'' state
\EndProcedure

\end{algorithmic}
\end{algorithm}

\subsection{Hybrid Monte Carlo simulation}
MCMC methods have many applications in the geosciences, in particular in Bayesian modeling where they enable the
numerical calculation of multiple integrals that cannot be evaluated by means of other methods~\citep{dth01,Norberg02,Caers06,Majumdar07}
However, standard MCMC with the Metropolis updating can be quite inefficient in the first stage of the simulation while the system relaxes towards the equilibrium distribution.
The latter is assumed to represent the true probability distribution of the system at the specified temperature.
The  inefficiency is due to the fact that the Metropolis updates involve
random proposals for the spins which are drawn from the uniform distribution $U[0, 2\pi]$. This  leads to small acceptance rates (most proposals are rejected).

To increase the efficiency of the relaxation procedure a hybrid MC algorithm (see Algorithm~\ref{algo:mpr-relax}) that combines a restricted form of the \emph{stochastic Metropolis}~\citep{metro87} and the \emph{deterministic over-relaxation}~\citep{creutz87} methods is implemented. The former algorithm generates a proposal spin-angle state at the \emph{i}th site according to the rule $\phi_i'=\phi_i+\alpha(r-0.5)$, where $r$ is a uniformly distributed random number $r \in [0,1)$ and $\alpha=2\pi/a \in (0,2\pi)$ is a tunable parameter automatically reset during the equilibration to maintain the acceptance ratio $A$ close to a target value $A_{\textrm{targ}}$. In the energy conserving over-relaxation update, new spin angle value at the \emph{i}th site is obtained by a simple reflection of the spin about its local molecular field, generated by its nearest neighbors, by the following transformation
\begin{equation}
\label{eq:over-relax}
\phi'_{i}   = \, \left[2\, \arctan2 \left( \sum_{j \in nn(i)} \sin{\phi_{j}}, \sum_{j \in nn(i)} \cos{\phi_{j}} \right)-\phi_{i}\right] \mod {2\pi},
\end{equation}
where $nn(i)$ denotes the nearest neighbors of $\phi_{i}$, $i=1, \ldots, N$, and $\arctan2(\cdot)$ is the four-quadrant inverse tangent: for any  $x, y \in \mathbb{R}$
such that $|x| + |y| >0$, $\arctan2(y, x)$ is the angle (in radians) between the positive horizontal axis and the point  $(x, y)$.

\begin{algorithm}[t!]
\caption{Simulation  of \mpr model. The algorithm involves the hybrid updating procedure \textsc{Update} described in Algorithm~\ref{algo:mpr-relax}.
$\mathbf{\Phi}_{s}$ is the vector of known spin values at the sample sites. $\mathbf{\hat{\Phi}}$  represents the vector of estimated spin values at the prediction sites.
$\hat{T}$ is the estimated reduced temperature. $G(\cdot)$ is the transformation from the original field to the spin field and $G^{-1}(\cdot)$ is its inverse. $\mathbf{\hat{Z}}(j)$, $j=1, \ldots, M$ is the $j$-th realization of the original field. $\mathbf{U}(0,2\pi)$ denotes a vector of random numbers from the uniform probability distribution in $[0, 2\pi]$. }
\label{algo:mpr-simul}
\begin{algorithmic}
\State 1: \emph{Set fixed simulation parameters}
\State 1.1: Set the following parameters (their effect on prediction performance was found to be marginal and the default values set by \citet{mz-dth18} can be used in general): $M$ - $\#$ equilibrium configurations for statistics collection, $n_{f}$ - verification frequency of equilibrium conditions, $n_{\textrm{fit}}$ - $\#$ fitting points of energy evolution function, $A_{\textrm{targ}}$ - target acceptance ratio of Metropolis update, $k_a$ - defines variation rate of perturbation control factor $a$, $i_{\max}$ - MC relaxation-to-equilibrium steps (optional).
\State 2: \emph{Initialize variable simulation parameters}
\State  2.1:  $ i \gets 0$      \Comment Initialize simulated state counter
\State  2.2:  $\mathbf{\hat{\Phi}}(0) \gets \mathbf{U}(0,2\pi)$  \Comment Initialize missing spin values from  uniform distribution
\State  2.3:  $ k(0) \gets - 1$    \Comment Initialize slope of energy evolution function
\State  2.4:  $ a(0) \gets 1$      \Comment Set spin angle perturbation control factor

\State 3: \emph{Data transformation}
\State  3.1: $\mathbf{\Phi}_{s} \gets G(\mathbf{Z}_{s})$  using~\eqref{map} \Comment Set data spin angles

\State 4: \emph{Parameter Inference}
\State 4.1: Estimate $e_s$ using~\eqref{eq:mpr-sse} \Comment Find sample specific energy
\State 4.2: $\hat{T} \gets e^{-1}(e_s|L)$ \Comment Estimate reduced temperature based on $e(\hat{T},L)=e_s$

\State 5: \emph{Non-equilibrium spin relaxation procedure}
\While{$[k(i)<0] \wedge [i \le i_{\max}]$ }    \Comment Spin updating with hybrid step
\State 5.1: \Call{Update}{$\mathbf{\hat{\Phi}}(i+1), \mathbf{\hat{\Phi}}(i), a(i),\hat{T}$}
\If{$A < A_{\textrm{targ}}$} \Comment Check if Metropolis acceptance ratio is low
\State 5.2: $a(i+1) \gets 1+(i+1)/k_a$ \Comment Update perturbation control factor
\EndIf
\State 5.3: Calculate $e(i+1) \gets \mathcal{H}/N_{GP}$ \Comment Obtain current specific energy
\If{$[i \geq n_{\textrm{fit}}] \wedge [i \equiv 0 \pmod{n_f}]$} \Comment Check frequency for slope update of $e$
\State 5.4:  $k(i+1) \gets \mathrm{SG}$ \Comment Update slope of $e$ by SG filter using last $n_{\textrm{fit}}$ values
\EndIf
\State 5.5: $ i \gets i +1$ \Comment Update MC counter
\EndWhile

\State 6. \emph{Equilibrium state simulation}
\State 6.1: $\mathbf{\hat{\Phi}}^{\mathrm{eq}}(0) \gets \mathbf{\hat{\Phi}}(i)$ \Comment Initialize the equilibrium state
\For{$j=0, \ldots, M-1$}
\State 6.2: \Call{Update}{$\mathbf{\hat{\Phi}}^{\mathrm{eq}}(j+1), \mathbf{\hat{\Phi}}^{\mathrm{eq}}(j), 1,\hat{T}$} \Comment Generate equilibrium realizations
\State 6.3: $\mathbf{\hat{Z}}(j+1)  \gets G^{-1}\left[\mathbf{\hat{\Phi}}^{\mathrm{eq}}(j+1)\right]$ \Comment Back-transform spin states
\EndFor
\State 7: \Return Statistics of $M$ realizations $\mathbf{\hat{Z}}(j), \, j=1, \ldots, M$

\end{algorithmic}
\end{algorithm}

A vectorized hybrid MC simulation algorithm (see Algorithm~\ref{algo:mpr-simul}) leads to fast relaxation followed by equilibrium simulation. The crossover to equilibrium (flat regime in the energy evolution curve) is automatically detected by periodic evaluation every $n_f$ MC sweeps applying the variable-degree polynomial Savitzky-Golay (SG) filter~\citep{savgol64} as the point where the trend disappears and the energy shows only fluctuations around a stable level. The total computational time scales approximately linearly with the system size. The demonstrated efficiency of the \mpr method makes it suitable for big data sets, such as satellite and radar images. The \mpr algorithm is described in detail in~\citep{mz-dth18}.

\section{CUDA Programming Model and its Implementation}
\label{sec:cuda}
CUDA is a general purpose parallel computing platform and programming model created by NVIDIA and implemented by the NVIDIA GPUs, which leverages the power of parallel computing on GPUs to solve complex computational problems more efficiently  than on a single CPU. CUDA comes with a software environment that extends the capabilities of C programing language, thus allowing developers to create their own parallelized applications. More details of the CUDA architecture are presented in~\citet{nvidia_cuda_2018}. Below, a brief overview of the CUDA programming model is  given and the features that are critical for the improved performance of the \mpr method are highlighted.

CUDA C code is essentially C code with additional syntax, keywords and API functions. Based on those the CUDA compiler (\textit{nvcc}) splits the code into multiple parts, which are executed on either the CPU, which is known as the \textit{host} (consequently, this part of the code is called \textit{host code}), or the GPU, which is known as the \textit{device} (hence, this part is known as the \textit{device code}). The most basic concept of a CUDA program is as follows: Some data is stored in the \textit{host} memory (i.e., RAM) representing input data for computations. First, the \textit{host} copies the data to the memory allocated on the \textit{device} and then the \textit{device} performs computations using \textit{device} functions called kernels. The latter harness the parallel computing capabilities of the GPU, solving the problem in an efficient and optimized manner. Finally, the \textit{device} copies the results back to the \textit{host} memory. In the following paragraphs the implementation of parallel computing by programming kernels is reviewed and the differences between various types of memory available on the GPU are explained. The proper setup of the CUDA programming model is crucial for optimal computational performance.

\subsection{CUDA Kernels}

A GPU kernel is a C/C++ function written in a serial way that can be executed on many GPU threads in parallel. A GPU kernel is run on blocks of threads organized into a grid\footnote{Note that the term grid in this subsection refers to a unit in the GPU architecture and should not be confused with the spatial grids used in the remainder of the paper.}. Threads are very simple software units --virtual processors-- that can be mapped to hardware resources to execute computations. The grid blocks and the number of threads per block are specified at  launch. Each thread can execute the same kernel function on different input data points. This approach is called SIMD (Single Instruction Multiple Data)  and implements data-level parallelism. Threads in a block as well as blocks in a grid can be arranged in one, two or three dimensions. The hardware limits the number of threads per block to a maximum of 1,024 (512 on older devices with compute capability\footnote{The compute capability of a GPU determines its general specifications and available features i.e maximum size of a block, how many blocks fit on a multiprocessor or whether or not is the GPU capable of certain operations.} less than 2.0). The blocks of threads are scheduled to run on the GPU cores organized into streaming multiprocessors (SMs). A single SM can execute one or more thread blocks. Individual threads are executed in groups called \textit{warps}. Warps get scheduled to run on an SM automatically by the warp scheduler, which can also stall warps waiting for data or memory transactions and schedule others in their place, so that as many warps as possible run concurrently. If SMs are occupied with a sufficiently large number of warps, this scheduler mechanism can effectively hide the high latency of memory operations, which is crucial for a highly performing CUDA application.

\subsection{Memory Hierarchy}
There are several types of memory available on the GPU, each with its own uses, advantages, and disadvantages. The three most important ones are reviewed below.

\subsubsection{Global Memory}
Global memory (also called device memory or DRAM) is located on the graphics card but off the GPU chip itself. Global memory can be accessed by any active thread in the entire grid; thus, it can be used for communication between any given threads. It is by far the largest (up to 48 GB on the latest professional cards, 4 GB on GTX 980 which is used in this work) but also the slowest, as it takes about 500 clock cycles to read or write data, compared to one clock cycle for an arithmetic operation. This is, however, still orders of magnitude faster than writing or reading directly from the host memory. Consequently, for performance reasons it is vital to maximize arithmetic intensity, that is, the number of arithmetic operations per read/write cycle. Another vital optimization strategy is called coalesced memory access, which is vital for both reading from and writing to the global memory ~\citep{harris_how_2013}.

\subsubsection{Shared Memory}
Shared memory, also called L1 cache, is located on the GPU chip  and provides an option for low latency communication between threads within a block, since reading or writing to it is much faster than using the global memory. The size of shared memory is, however, severely limited, for instance at 48 KB per thread block and 96 KB per SM on our GTX 980. A common strategy for utilizing shared memory is as follows: First, store the data needed for the block in its shared memory, then perform computations requiring reading/writing only on the shared data, and finally output the block results to the global memory for further use. This approach minimizes high latency global memory access and it is desirable in reduction algorithms, such as the calculation of a sum of data array.

\subsubsection{Registers}
The per thread local memory is implemented by 32 bit (4 byte) registers and is used for storing local variables of a particular thread. Registers provide extremely low latency memory access. In past GPU architectures (before Kepler) this per thread local memory was only accessible by the same thread. However, with the advent of the Kepler architecture (2012), threads belonging to the same warp can read each other's registers using a warp shuffle operation. The number of registers used per block is capped to a maximum of 65,536.

\subsubsection{Other types of memory}
For completeness, there are two more types of memory --texture and constant. They both reside in DRAM, are read-only and cached. If properly used, they should improve performance over global memory reads. The constant memory is optimized for broadcasting the same value to all threads in warp, while the texture memory is more suitable for data with 2D spatial locality.

\subsection{Implementation of the \mpr Method}
Achieving maximum speedup of the \mpr method hinges on efficient (parallel) implementation of the hybrid algorithm that combines deterministic over-relaxation with a stochastic Metropolis step. Our approach is based on the parallel implementation of the Metropolis algorithm for the Ising model~\citep{weig11,weig12}.

\begin{figure}[t]
\centering
  \includegraphics[scale=0.25,clip]{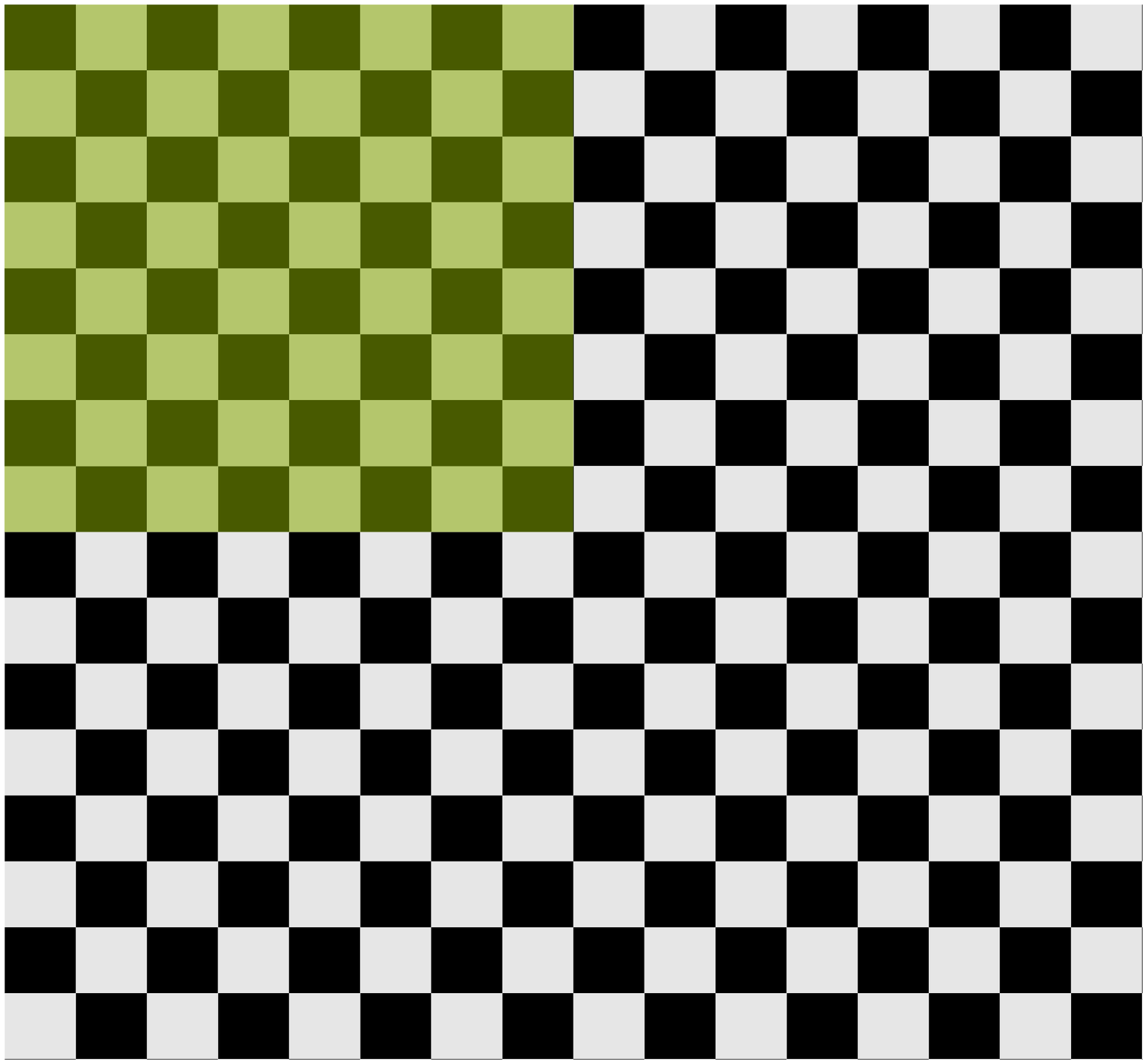}
  \caption{Checkerboard decomposition of the spin grid. Each square represents a grid node with its associated spin variable. Sub-grid A consists of the dark nodes, while sub-grid B comprises the light nodes. The green (shaded) tile in the upper left corner represents a block of threads that are responsible for the numerical operations on the spins included within the square. Each thread performs calculations for a single spin.}
  \label{fig:checkerboard}
\end{figure}

In the following it is assumed that the data are supported on a square, two-dimensional grid (extension to rectangular grids is straightforward). Square grids are bipartite graphs, meaning that they can be divided into two disjoint and independent graphs (sub-grids), for example, A and B. Most importantly, the nearest neighbors of any node on the sub-grid A belong to the sub-grid B, and vice versa. Hence, the \mpr model's set of variables (``spins'') can be split between two sub-grids, so that spins in the first sub-grid only interact with  spins of the second sub-grid and vice versa. This is called the checkerboard decomposition, and it is depicted in Fig.~\ref{fig:checkerboard}. By means of this decomposition it is possible to apply the updating algorithm to all the sub-grid spins in parallel. To perform the necessary calculations, one thread is called per each  sub-grid spin.

Let us recall that the used hybrid MC updating scheme involves two steps: the deterministic over-relaxation update that conserves the energy is combined with the stochastic Metropolis update to achieve ergodicity. In the case of the over-relaxation step, each thread calculates the local energy contribution of its assigned spin on the first sub-grid, and subsequently updates the spin value so that the local energy contribution remains unchanged. In the case of the Metropolis step, each thread loads both the old and new spin states into its registers, calculates the local energy difference between the states, and attempts to update the spin according to the Metropolis criterion. Once these steps are completed, the same procedure is also applied to the second sub-grid.

\begin{figure}[t]
\centering
  \includegraphics[scale=0.35,clip]{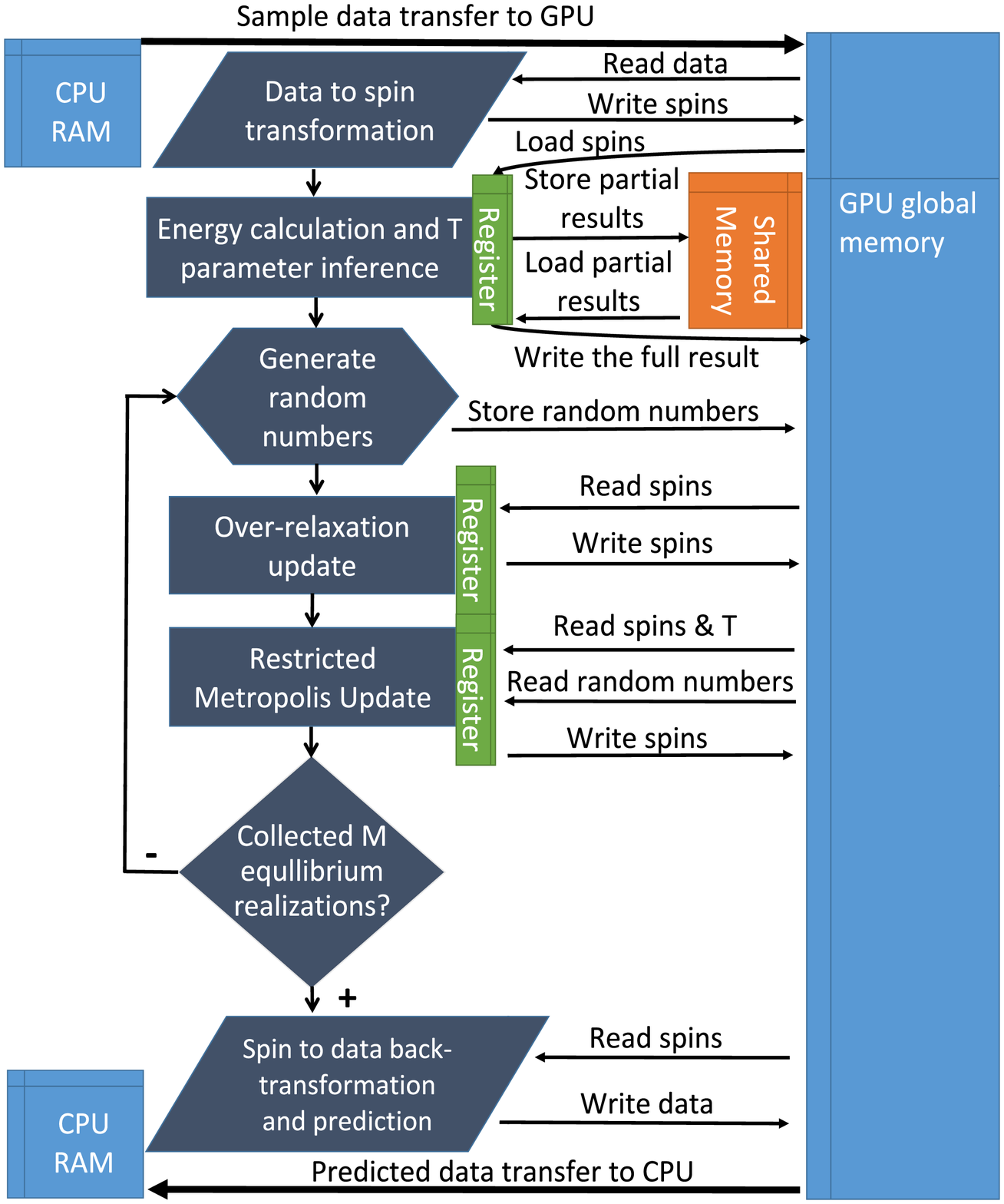}
  \caption{Representation of the main computational steps and memory transactions performed on the GPU.}
  \label{fig:flowchart}
\end{figure}

To avoid any unnecessary overhead, it is important to generate the random numbers needed for proposing new states and the Metropolis updates beforehand and store them in the GPU’s global memory. For each spin and each MC step two random numbers are needed --- one to propose a new state for the given spin and another one to stochastically accept it. They were generated by \verb+Philox4_32_10+ random number generator from \verb+cuRAND+ library. A flowchart depicting the main steps of GPU implementation is shown in Fig.~\ref{fig:flowchart}.

\subsubsection{Launch Configuration}
The kernel launch configuration  for the spin updating procedure is an important aspect of the implementation. It involves various settings, such as specifying grid and block dimensions and the amount of shared memory that the kernel expects to use. As many threads are needed as there are spins in each sub-grid arranged into a grid of thread blocks in a way that keeps the index arithmetic as simple as possible, yet also coalesces global memory access.

For our two-dimensional (2D) system it is natural to use a 2D grid of 2D blocks (squares). The block size, however, should be carefully considered as it can affect hardware utilization, which is measured by a metric called occupancy~\citep{nvidia_cuda_2018-1}. Occupancy is the ratio of the number of active warps per SM to the maximum number of possible active warps. Two factors limit how many warps can be run on a SM in parallel: the maximum number of blocks per SM, which depends on the compute capability of the device and can be found in its technical specifications, and the availability of resources such as the shared memory and registers. The block size  is limited to 1,024 threads per block on devices with  at least 2.0 compute capability. The optimal size of a (square) block should always be a multiple of warp size (32) and, thus, considering the limitation of 1,024 threads per block, the candidate sizes are 64, 256, and 1,024. Performance based on different block sizes (parametrized by their side length $B$) is presented in Sect.~\ref{sec:results}.

\subsubsection{Effects of Arithmetic Precision}
The numerical precision of the calculations affects the execution time on the GPU considerably more than on the CPU. If the problem permits, one can achieve remarkable speedups by switching from double to single precision, and by making use of fast, precision-specific CUDA math functions. For example, if we opt for single precision, instead of functions such as \texttt{sin(), cos() or exp()} we can use their single precision counterparts, that is,  \texttt{sinf(), cosf() and expf()}, which are significantly faster than their more precise analogues. However, it should be ensured that using lower precision does not severely affect the results due to excessive rounding errors.

\section{Results}
\label{sec:results}
The performance of the \mpr method, implemented on CPU in the C++ and on GPU in the CUDA environments, is validated using synthetically generated spatial data. For the sake of consistency, data with the same statistical properties as in~\citet{mz-dth18} are generated; therein the data were processed on CPU in the Matlab\textregistered\ environment. Namely, samples from the Gaussian random field $Z \sim N(m=50,\sigma=10)$  with Whittle-Mat\'{e}rn (WM) covariance
are simulated on square grids with $L$ nodes per side for different values of $L$. The covariance function is given by
\begin{equation}
\label{mate} G_{\rm Z} (\|\mathbf{h}\|)=
\frac{{2}^{1-\nu}\, \sigma^{2}}{\Gamma(\nu)}(\kappa \,
\|\mathbf{h}\|)^{\nu}K_{\nu}(\kappa \, \|\mathbf{h}\|),
\end{equation}
where $\|\mathbf{h}\|$ is the Euclidean two-point distance, $\sigma^2$
is the variance, $\nu$ is the smoothness parameter, $\kappa$ is the
inverse autocorrelation length, and $K_{\nu}$ is the modified Bessel
function of index $\nu$.  The simulations are based on the spectral method~\citep{drum87}

From the simulated data, $S=100$ different sampling configurations are generated by random removal of $p=33\%$ and $66\%$ of points. The \mpr predictions at the removed points are calculated and compared with the true values using different validation measures. For consistency with the Matlab\textregistered\ results reported by~\citet{mz-dth18},  the same validation measures are evaluated, that is, the mean average absolute error (MAAE), the mean average relative error (MARE), the mean average absolute relative error (MAARE), and the mean root average squared error (MRASE). For both the CPU and the GPU calculations   the respective run times $\langle t_{\mathrm{cpu}} \rangle$ and $\langle t_{\mathrm{gpu}} \rangle$ are measured and the speedup of the CPU Matlab\textregistered\ as well as GPU CUDA implementations relative to the single-CPU C++ implementation is evaluated. The calculations were conducted on a desktop computer with 16.0 GB RAM, Intel\textregistered Core\texttrademark2 i5-4440 CPU processor and NVIDIA GeForce GTX 980 GPU. The computer's technical specifications are summarized in  Table~\ref{tab:t_tsp}.

\begin{table}
\caption{Comparison of the main technical specifications of  hardware used in the numerical experiments:  Intel\textregistered Core\texttrademark2 i5-4440 CPU and  NVIDIA GeForce GTX 980 GPU.  Note that 1 GFLOP = $10^{9}$ floating point operations per second and 1 TFLOP =  $10^{12}$ floating point operations per second. MSVC: Microsoft Visual C++ Compiler. NVCC: NVIDIA CUDA Compiler.}
\label{tab:t_tsp}       
\begin{tabular}{lcc}
\hline\noalign{\smallskip}

          & CPU - Intel Core i5-4440& GPU - NVIDIA GTX 980 \\
					\hline
    $\#$ of Cores &  4 & 2,048\\
    Clock Frequency &  3.10 GHz & 1.127 GHz\\
    Cache  &  6 MB & 768KB L1 + 2MB L2 \\
    Theoretical peak performance (SP) &  396.8 GFLOPS & 4.981 TFLOPS\\
    Max memory bandwidth&  25.6 GB/s & 224.4 GB/s\\
    Compiler &  MSVC & NVCC\\
\noalign{\smallskip}\hline
\end{tabular}
\end{table}

In Table~\ref{tab:t_dp} validation measures obtained with double-precision (DP) arithmetic are presented for random fields with WM covariance parameters $\nu=0.5,\kappa=0.2$ on a square grid with  $L = 1,024$ in the three different programming environments. There are only minute differences between the validation measures obtained by the Matlab\textregistered\ , C++ and CUDA implementations, which are most likely caused by using different sequences of random numbers. Similar results are also obtained for other values of the WM parameters.  Therefore, in the following we set $\nu=0.5,\kappa=0.2$ and focus on other factors that  affect computational efficiency. In particular, the choice of the programming environment can strongly affect the computational efficiency. The bottom row of Table~\ref{tab:t_dp} shows that, in spite of the vectorization based on the checkerboard algorithm the Matlab\textregistered\ code is about $7-9\%$ slower than the C++ code. This is not surprising, as the former is well known to be less efficient, particularly if the algorithm involves a large number of iterations. On the other hand, the GPU implementation leads to a dramatic increase of speed. More specifically, for the chosen block side length $B = 16$ and  data sparsity $p=33\%$, the CUDA GPU time is 43 times smaller than the C++ CPU time and the speedup increases with $p$ up to almost 80 times for $p=66\%$.

\begin{table}
\caption{Validation measures obtained in different programming environments for Gaussian random field data with  WM covariance parameters $\nu=0.5,\kappa=0.2$. The data are generated on a square grid with length $L = 1,024$ per side for different values of sparsity, and the calculations use \emph{double precision} arithmetic.}
\label{tab:t_dp}       
\begin{tabular}{lcccccc}
\hline\noalign{\smallskip}
          & \multicolumn{2}{c}{CPU (C++)} & \multicolumn{2}{c}{CPU (MATLAB)} & \multicolumn{2}{c}{GPU ($B = 16$)} \\
          & $p = 33\%$ & $p = 66\%$ & $p = 33\%$ & $p = 66\%$ & $p = 33\%$ & $p = 66\%$ \\
					\hline
    MAAE  & 3.381 & 3.828 & 3.379 & 3.828 & 3.380 & 3.827 \\
    MARE [\%] & $-$0.989 & $-$1.303 & $-$0.993 & $-$1.299 & $-$0.991 & $-$1.297 \\
    MAARE [\%] & 7.172 & 8.157 & 7.171 & 8.157 & 7.172 & 8.156 \\
    MRASE & 4.245 & 4.821 & 4.247 & 4.822 & 4.244 & 4.821 \\
    $\langle t^{dp} \rangle$ [s] & 38.665 & 69.953 & 42.484 & 75.092 & 0.893 & 0.883 \\
    speedup & 1.000 & 1.000 & 0.910 & 0.932 & 43.279 & 79.200 \\
\noalign{\smallskip}\hline
\end{tabular}
\end{table}

Further appreciable  gains in speed without noticeable compromise of the predictive performance can be obtained by opting for single-precision (SP) calculations. Table~\ref{tab:t_sp} shows that for the same parameters as in Table~\ref{tab:t_dp} the change to  single precision  increases the speedup by almost four times, that is, to 166 for $p=33\%$ and up to 306 for $p=66\%$, while all the validations measures remain the same (up to at least the third decimal place). Table~\ref{tab:t_sp} also demonstrates the effect of the block size selection. The value of the block length $B$ does not seem to affect the prediction performance, but it can be optimized with respect to speed. Based on our test results, optimal speed is achieved for $B = 16$. Choosing other block values (particularly smaller, i.e., $B = 8$) leads to less efficient performance. This is in agreement with the findings reported by~\citet{weig11,weig12}.

\begin{table}
\caption{Validation measures obtained in different programming environments for Gaussian random field data with  WM covariance parameters $\nu=0.5,\kappa=0.2$. The data are generated on a square grid with length $L = 1,024$ per side for different values of sparsity, and the calculations use \emph{single precision} arithmetic.}
\label{tab:t_sp}       
\begin{tabular}{lcccccccc}
\hline\noalign{\smallskip}
          & \multicolumn{2}{c}{CPU (C++)} & \multicolumn{2}{c}{GPU ($B = 8$)} & \multicolumn{2}{c}{GPU ($B = 16$)} & \multicolumn{2}{c}{GPU ($B = 32$)} \\
          & $p = 33\%$ & $p = 66\%$ & $p = 33\%$ & $p = 66\%$ & $p = 33\%$ & $p = 66\%$ & $p = 33\%$ & $p = 66\%$ \\
					\hline
    MAAE  & 3.380 & 3.828 & 3.380 & 3.827 & 3.380 & 3.827 & 3.380 & 3.827 \\
    MARE [\%] & $-$0.992 & $-$1.300 & $-$0.991 & $-$1.297 & $-$0.991 & $-$1.297 & $-$0.990 & $-$1.297 \\
    MAARE [\%] & 7.172 & 8.158 & 7.172 & 8.156 & 7.172 & 8.156 & 7.172 & 8.156 \\
    MRASE & 4.244 & 4.821 & 4.244 & 4.821 & 4.244 & 4.821 & 4.244 & 4.821 \\
    $\langle t^{sp} \rangle$ [s] & 26.899 & 52.771 & 0.207 & 0.223 & 0.162 & 0.173 & 0.164 & 0.174 \\
    speedup & 1.000 & 1.000 & 129.972 & 236.575 & 166.474 & 305.795 & 164.091 & 303.593 \\
\noalign{\smallskip}\hline
\end{tabular}
\end{table}

\begin{figure}[tpb]
		\subfigure{\label{fig:t_dp}\includegraphics[scale=0.4,clip]{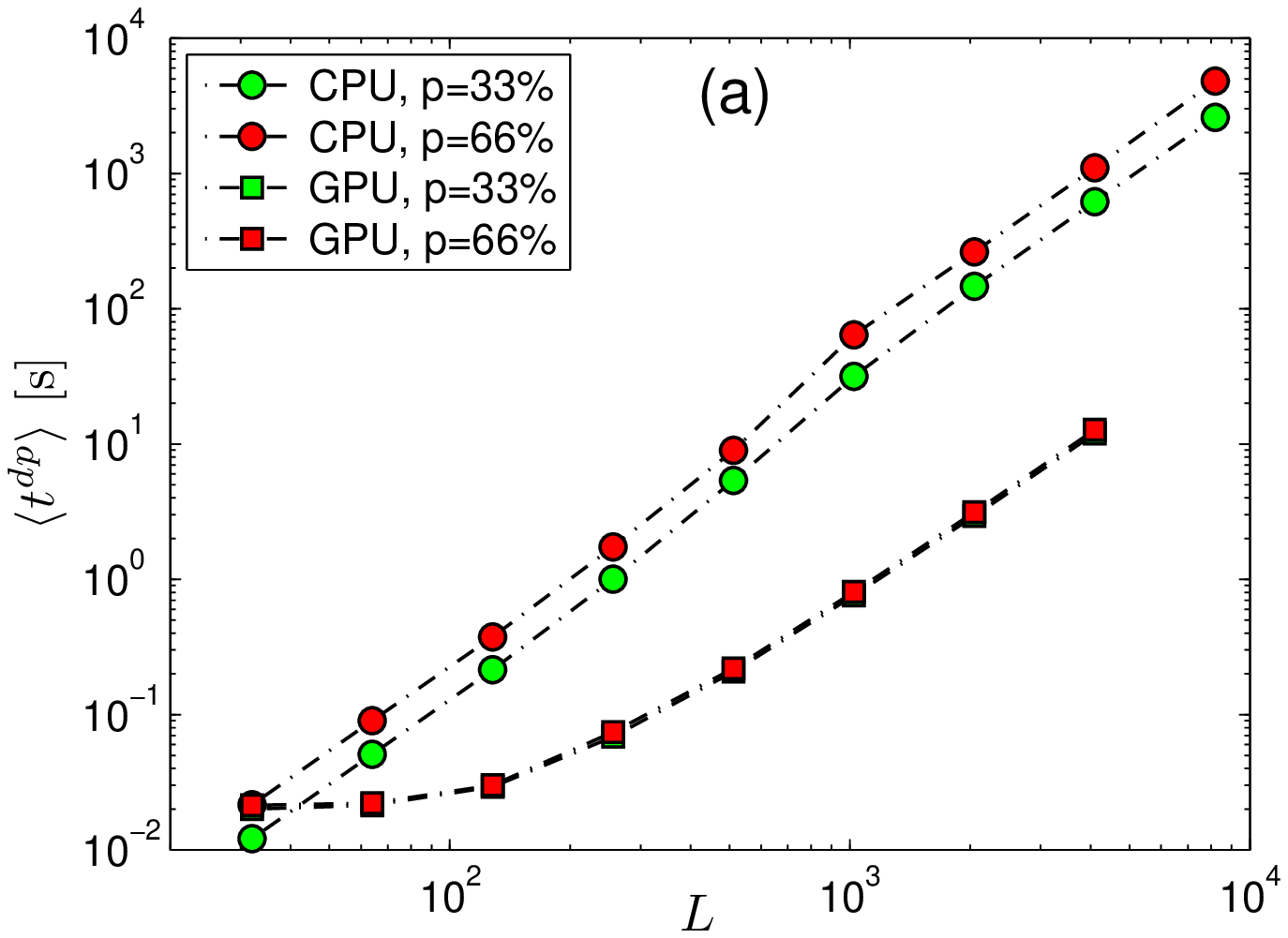}}
    \subfigure{\label{fig:t_sp}\includegraphics[scale=0.4,clip]{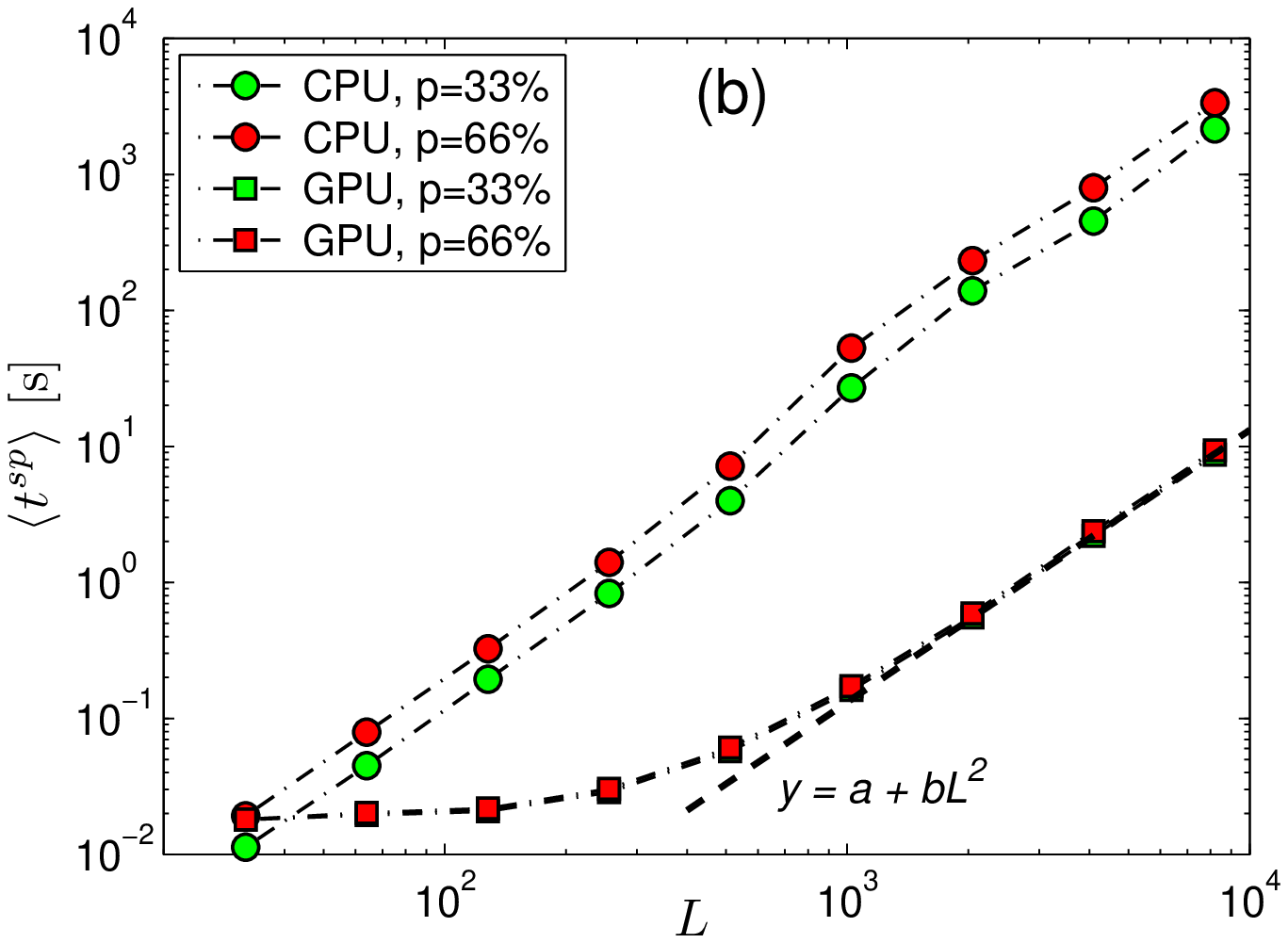}}
    \caption{Comparison of the computational times (in seconds) required by CPU (circles) and GPU (squares) calculations and their dependence on the grid length $L$, for $p=33\%$ (green) and $p=66\%$ (red). Panels (a) and (b) show results for  double and single precision arithmetic, respectively. The dashed line in (b) shows the linear fit to the grid size $L^2$. The GPU times are little sensitive to the data sparsity (green and red squares almost coincide).}
  \label{fig:t}
\end{figure}

\begin{figure}[tpb]
\centering
\includegraphics[scale=0.7,clip]{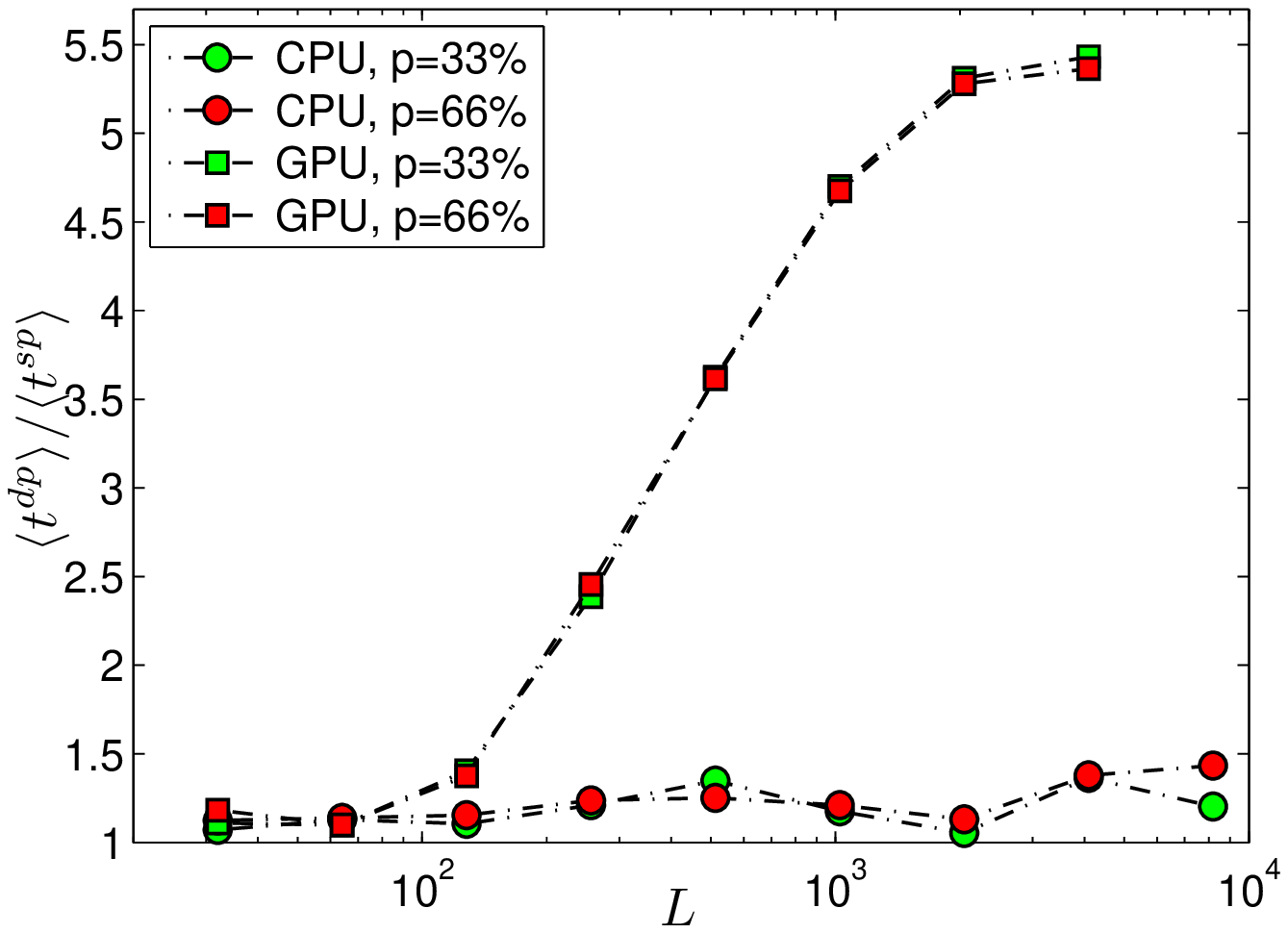}
\caption{Ratio of double- over the single-precision computational times obtained on CPU (circles) and GPU (squares), for $p=33\%$ (green) and $66\%$ (red) versus the grid length $L$.  }
  \label{fig:t_dp-sp}
\end{figure}

The results presented in the above tables pertain to fixed grid side length $L=1,024$. Targeting massive data sets, it is interesting to investigate the dependence of computational time on increasing grid size. Figure~\ref{fig:t} illustrates the dependence of the computational time, on both CPU and GPU, as well as for different sparsity values $p$. Figs.~\ref{fig:t_dp} and~\ref{fig:t_sp} show results for the double- and single-precision calculations, respectively. For the CPU data, the increase of the computational time follows an approximately linear dependence at all scales, in line with the Matlab\textregistered\ checkerboard results in~\citet{mz-dth18}. The CPU times range between a few milliseconds for the smallest, $L=32$, up to about one hour for the largest, $L=8,192$ grid lengths. Naturally, the single-precision calculations are faster than the respective double-precision ones. However, the ratio of the computational times does not seem to vary systematically with $L$ and  does not exceed $44\%$.

On the other hand, there appear to be two regimes for the GPU times. For small to moderate $L$ ($L \sim 10^2 - 10^3$) the increase with size is very gentle due to the low SMs occupancy that cannot completely hide memory latency. Only for larger grid sizes, when SMs are fully utilized, the GPU times follow about the same linear increase as the CPU ones (see the dashed line in Fig.~\ref{fig:t_sp}). In contrast with the CPU calculations, the relative difference between the SP-GPU and DP-GPU times systematically increases for $L > 68$;  for $L$ exceeding 1,024 the SP-GPU execution is faster than the DP-GPU, even by as much as five times, as shown in Fig.~\ref{fig:t_dp-sp}. Consequently, the SP-GPU times range between a few milliseconds for the smallest $L=32$ (for small $L$ they are even slightly larger than the CPU times) up to about nine seconds for the largest $L=8,192$. Note that the DP-GPU calculations could not be performed for the largest, $L=8,192$, grid length due to insufficient global memory.

\begin{figure}[t!]
		\subfigure{\label{fig:cpu-p}\includegraphics[scale=0.4,clip]{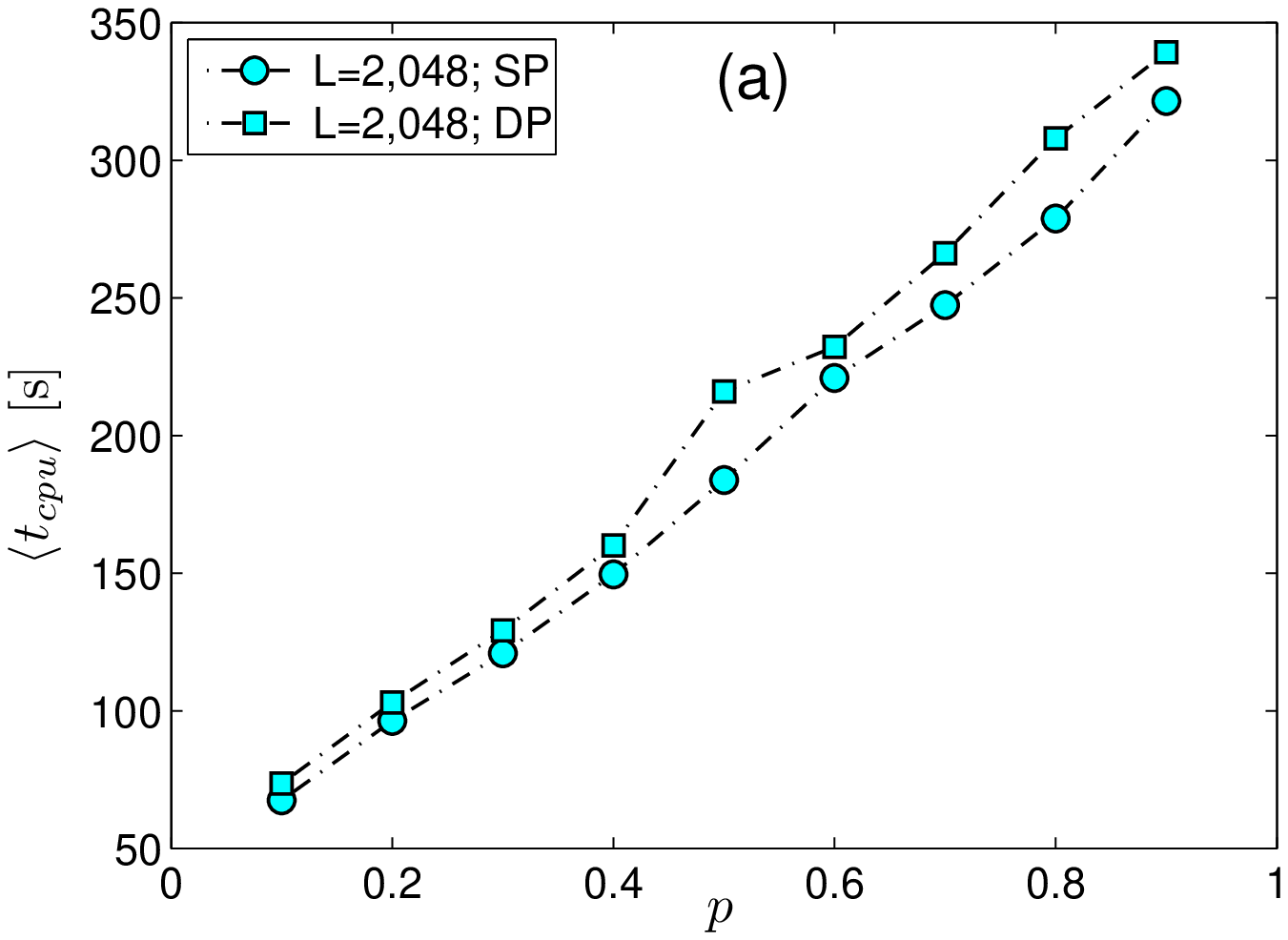}}
    \subfigure{\label{fig:gpu-p}\includegraphics[scale=0.4,clip]{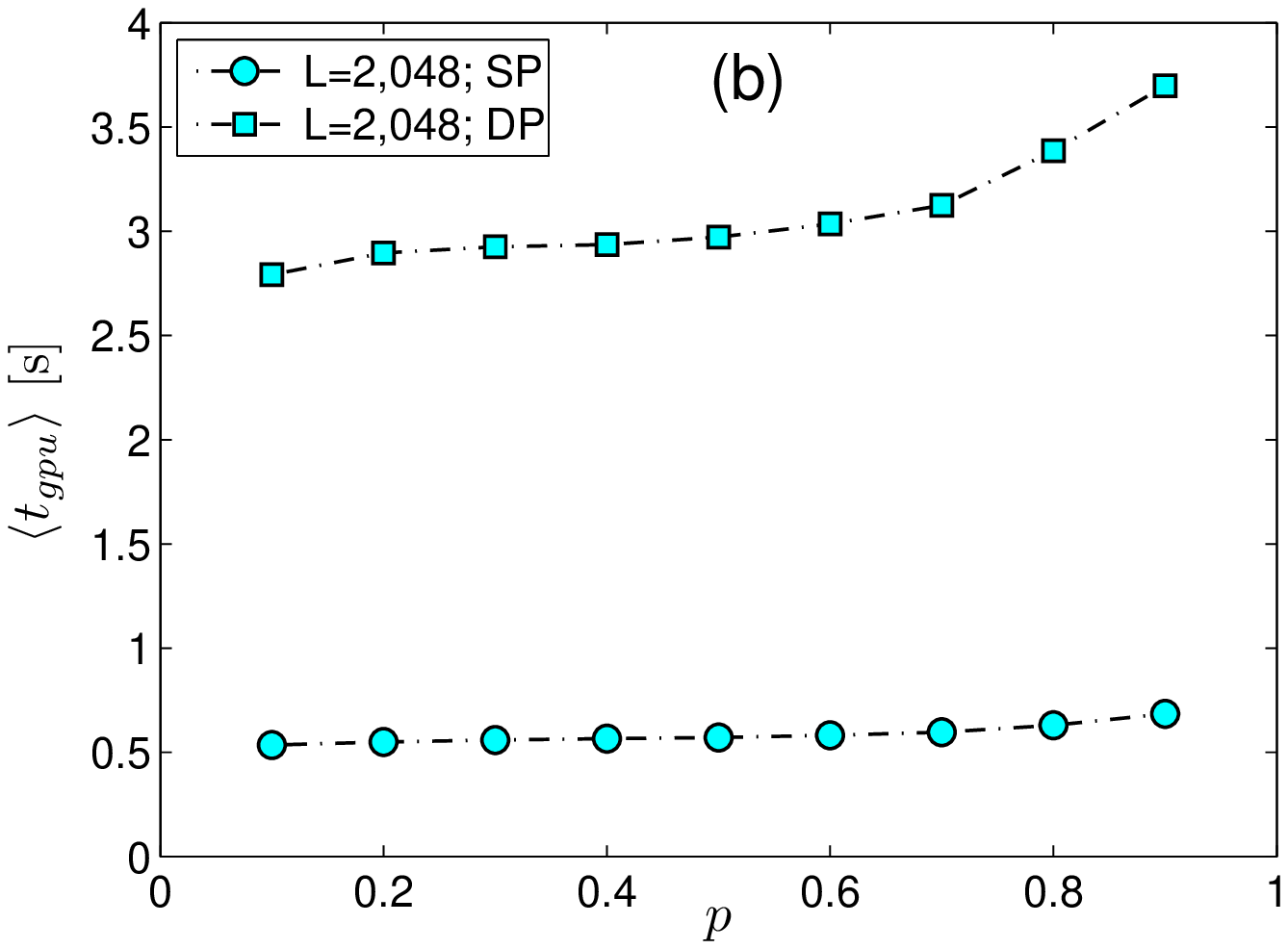}}
    \caption{(a) CPU and (b) GPU times as functions of the sample sparsity $p$, obtained by DP (circles) and SP (squares) calculations. The grid length per side is $L=2,048$.}
  \label{fig:t-p}
\end{figure}

Finally, the impact of the sample's sparsity on computational speed is investigated. Generally, higher sparsity means a higher number of prediction points that enter conditional Monte Carlo simulations, and therefore larger computational demands. In Fig.~\ref{fig:cpu-p} it is shown that the CPU time increases roughly linearly with sparsity $p$, for both DP and SP calculations. On the other hand, the GPU times presented in Fig.~\ref{fig:gpu-p} display a more complex behavior with  increasing $p$. Namely, a steeper increase for smaller $p$ is followed by a flatter part within $20\% \lesssim p \lesssim 70\%$ which is then followed by another steep increase for $p \gtrsim 70\%$. Again, the ratio of the DP-GPU over the SP-GPU times, that is,
$\langle t_{gpu}^{dp} \rangle/\langle t_{gpu}^{sp} \rangle \approx 5.1$,  significantly exceeds the value of the ratio, that is, $\langle t_{cpu}^{dp} \rangle/\langle t_{cpu}^{sp} \rangle \approx 1.1$, for the CPU calculations.

\begin{figure}[t!]
     \subfigure{\label{fig:speedup-L}\includegraphics[scale=0.4,clip]{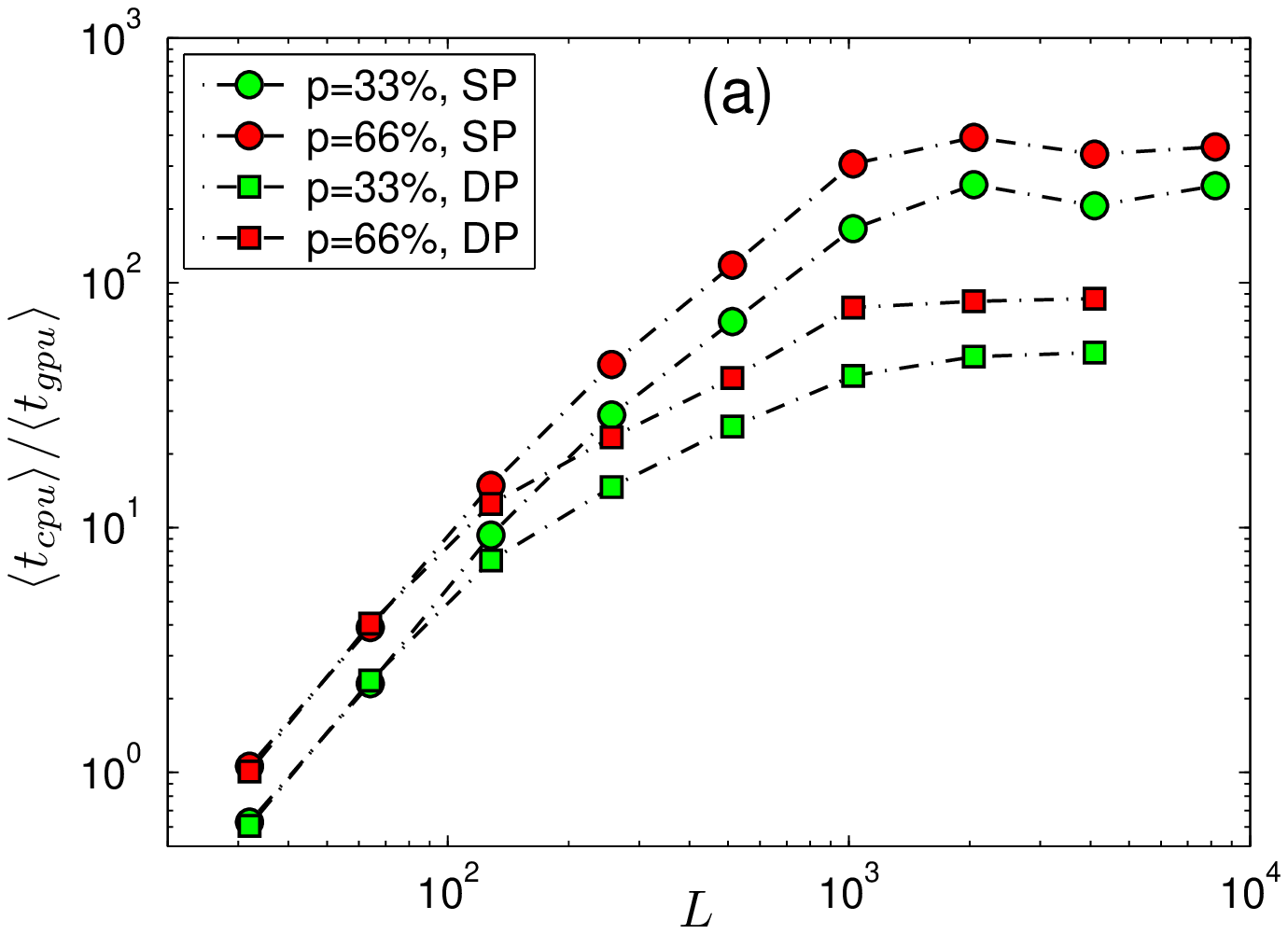}}
		 \subfigure{\label{fig:speedup-p}\includegraphics[scale=0.4,clip]{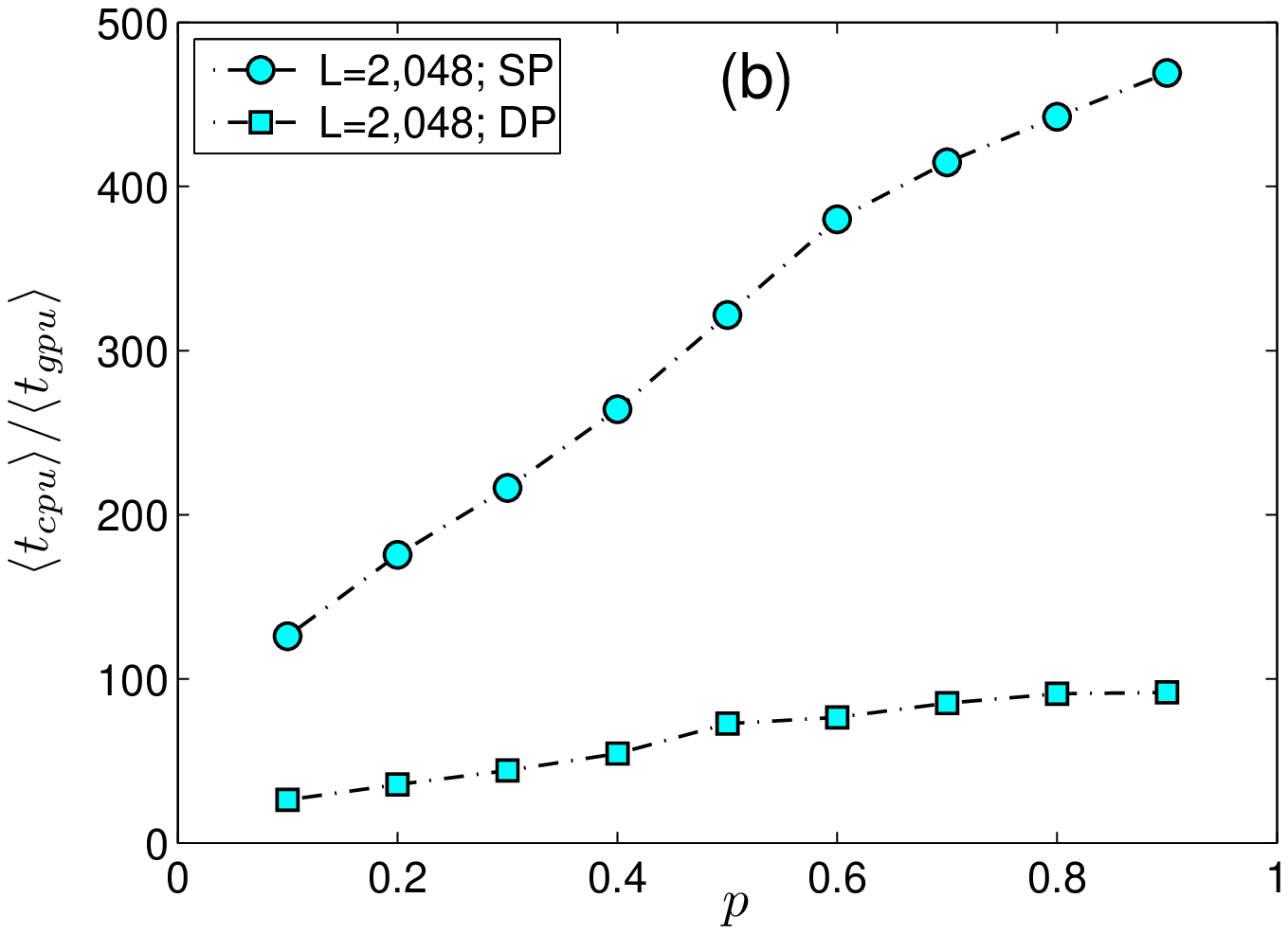}}
    \caption{Speedups of the GPU-accelerated \mpr method as functions of (a) the grid length $L$ for data with $p=33\%$ (green) and $p=66\%$ (red) and (b) the sparsity $p$ for $L=2,048$, obtained from SP (circles) and DP (squares) calculations.}
  \label{fig:speedup}
\end{figure}

Figure~\ref{fig:speedup-L} shows the speedup achieved by the GPU CUDA implementation compared to the single-CPU C++ implementation as a function of the grid length, for SP (circles) and DP (squares) calculations and two values of the thinning $p=33\%$ (green symbols) and $66\%$ (red symbols). As is evident, for the smallest grid length  $L=32$ the GPU implementation has no advantage over the CPU implementation in terms of computational speed. In fact, for $p=33\%$ the GPU run time is even larger than the CPU time (speedup is less then one). Nevertheless, the speedup factor dramatically increases with the grid size, and the full potential of the GPU code shows up for lengths $L \gtrsim 1,024$, at which all the speedup curves appear to level off. The speedup achieved by SP  calculations increases with the  grid length faster than the speedup for DP calculations. For example, for $L=32$ the speedup of the DP calculations is 0.61 for $p=33\%$ and 1.01 for $p=66\%$ versus 0.63 for $p=33\%$ and 1.06 for $p=66\%$ recorded for the SP calculations. On the other hand, for $L=2,048$ the DP values are 49.92 for $p=33\%$ and 83.95 for $p=66\%$ versus the SP values of 251.50 for $p=33\%$ and 392.10 for $p=66\%$.

The speedup increase with the sample sparsity $p$ is illustrated in Fig.~\ref{fig:speedup-p}. The speedup increases for both the DP and SP schemes, but the magnitude of the latter is more than five times larger for every $p$. In particular, for very sparse samples the SP calculations on GPU can be almost 500 times faster than those performed on CPU.

\section{Summary and Conclusion}
\label{sec:conclusion}
The ever-increasing amount of spatial data calls for  new computationally efficient methods and for  optimal use of available computational resources. Recently,  a novel spatial prediction method (\mpr), inspired from statistical physics, was introduced for the reconstruction of missing data on regular grids. In spite of its simplicity, the \mpr method was shown to  competitive with several interpolation methods as well as computationally  efficient~\citep{mz-dth18}. The \mpr efficiency derives from the local nature of the interactions between the \mpr model's variables and from an efficient hybrid simulation algorithm. Thus, the computational time of the \mpr method scales approximately linearly with system size. The computational speed along with the ability for automatic operation make the \mpr method promising for near real-time processing of massive raster data. Further gains in efficiency can be achieved by  memory use optimization and the algorithm's parallelization.

In the present paper, advantage of the local (nearest-neighbor) interactions of the \mpr model is taken to provide a  parallel implementation on general purpose GPUs in the CUDA environment. To demonstrate the computational speedup achievable by the GPU implementation, tests on synthetic data sets with randomly missing values are performed. The data sets have different sizes and sparsity. The tests are run in both CUDA on GPU and C++ on a single CPU. It is shown that the speedup can be optimized by a thoughtful setting the GPU environment, such as the block size. For the range of grid sizes in this study, a block side length $B=16$ is found to be optimal.

In line with our earlier results~\citep{mz-dth18}, the CPU time is confirmed to increase approximately linearly with the grid size $L^2$ over the entire range of  $L$ studied. On the other hand, the increase of the GPU time with grid length is initially very gentle  until the linear regime is established for $L \gtrsim 2,048$. Another advantage of the GPU over the CPU implementation is opting for single- instead of double-precision calculations. While there is no significant gain in single over double precision on CPU, in the GPU implementation for large enough $L$ the speedup can be more than  fivefold with no observable deterioration of the prediction performance. The speedup is also found to increase with the sample sparsity. For very sparse data on large grids, the speedup due to single precision arithmetic can be as large as almost 500 times. Thus, using an ordinary personal computer, data sets with arbitrary sparsity that involve up to hundreds of thousands of points can be processed in almost real time, and data sets that involve millions of points can be processed in less than one second.

\begin{acknowledgements}
This work was supported by the Scientific Grant Agency of Ministry of Education of Slovak Republic (Grant No. 1/0531/19). We also acknowledge support for a short visit by M.~\v{Z}. at the Technical University of Crete from the Hellenic Ministry of Education - Department of Inter-University Relations, the State Scholarships Foundation of Greece and the Slovak Republic's Ministry of Education through the Bilateral Programme of Educational Exchanges between Greece and Slovakia.
\end{acknowledgements}

\bibliographystyle{MG}       
{\footnotesize

\end{document}